\documentstyle{article}

\begin{document}
\title{Detection of massive multi-particle beams by two-particle ionization}
\author{Pedro Sancho \\ GPV de Valladolid \\ Centro Zonal en Castilla y
Le\'on  \\ Ori\'on 1, 47014, Valladolid, Spain}
\maketitle
\begin{abstract}
Multi-photon absorption is a well-known phenomenon. With atom
lasers a similar process could take place for massive particles,
the ionization of an atom or molecule by the successive
interaction with various particles. This process would lead to
multi-particle detection events for incident multi-particle beams.
We show that two-particle detections would introduce a correction
(proportional to the fourth power of the wavefunction modulus) to
the usual one-particle detection probability (only proportional to
the second power).
\end{abstract}
\vspace{7mm}



\section{Introduction}
The advent of the laser, providing a source of light beams with
high degrees of coherence was a major advance in the understanding
and manipulation of the quantum nature of light. Similarly, the
recent arrival of atom lasers could open new conceptual and
experimental areas in the field of atom optics, the study of
quantum properties of matter waves \cite{Rol}. The atom lasers
provide sources of matter waves with many particles in the same
state. It emerges naturally the question if the usual theory of
detection will suffice to correctly describe measurements on
multi-particle beams or if new effects could be present. We
analyze here a new aspect of the problem, the possible existence
of two-particle detection processes (single detection events
produced by the interaction with two particles of the beam). At
this point we return to quantum optics.

One of the advances associated with the laser was the improvement
in the study of multi-photon absorption, a central issue in
non-linear optics. An atom can absorb in consecutive steps various
incident photons \cite{Del}. The beams used in two-photon
absorption can be obtained from different classical light sources.
However, the intensity is in general weak and the non-linear terms
can be neglected. Only, for light from a laser source the
non-linear terms are in general relevant. Having at our disposal
matter wave beams with various particles in the same state, we
can pose the question if similar properties could be observed for
particles obeying Schr\"{o}dinger's equation. Massive particles
can ionize atoms and molecules. One expects that when these atoms
and molecules interact with multi-particle beams there can be
processes where they are not ionized by a single particle, but by
successive interactions with the particles of the beam. This
property could have implications in the context of detection
theory. As many detectors of massive particles are based on the
ionization induced by the probe particles these processes  would
lead to {\it multi-detection events} (single detection events, for
instance single detection clicks or spots, produced by various
particles).

We shall show as the main result of the paper that the detection
probabilities expected in the presence of {\it multi-detection
events} are, in principle, distinguishable from those associated
with {\it single-detection events}, opening the door to a possible
experimental verification of these phenomena. To be concrete, the
first type of event is proportional for particles in the same
state to the fourth power of the modulus of the wavefunction,
whereas for the second type it is only proportional to the second
power. This bevaviour is similar to that found in quantum optics,
where the probabilities of single- and double-absorption are
proportional to the intensity and to the squared intensity. In
particular, the interference patterns for one- or two-particle
incident beams would be slightly different. The results obtained
here for detection by ionization would remain unchanged for any
type of detection scheme with {\it multi-detection events}.

\section{Multi-detection events}

Several common methods of detection for massive particles rest on
the ionizations these particles can induce on the atoms or
molecules composing the detector. Well-known examples are those of
plates (used for both massless and massive particles), Geiger counters or bubblechambers.

When the underlying physical mechanism in the detection process is the
ionization we can think along very similar lines to the
multi-photon absorption. In the case of a single incident
particle the interaction with the detector produces directly the
ionization (or fails to produce ionization). With two incident
particles, we can have other ways to the final ionization. The
first incident particle can carry out the atoms or molecules of the
detector to an excited state, but without ionization. Later, the
interaction with a second particle definitively ionizes the
detector. The similarity with the multiple absorption of photons
is clear.

Usually atom laser beams and in general multi-particle beams are
obtained from Bose-Einstein condensates \cite{Adw}. The components
of these condensates are neutral, whereas the effects of
ionization are most times associated with charged particles. To
convert the original beam into one with charged particles we can
use a laser which ionizes the components of the beam.

From now on we concentrate on the possible modifications of the
detection probabilities associated with {\it multi-detection
events}.

\section{Detection probabilities}

Let us consider a small size detector placed at a given
position, which is used to measure the probability of finding at that position
particles \cite{com}. The incident beam is in a multi-particle state.

The probabilities for single and double detection events at point
${\bf r}$ (the position of the detector) and time $t$, in spin states $\mu $ or $\eta
$, are given by the well-known expressions:
\begin{equation}
P_{sin}({\bf r},t; \mu  \vee \eta )= \sum _{\xi =\mu , \eta
}\frac{<I|\hat{\psi} ^+_{\xi } ({\bf r},t) \hat{\psi} _{\xi}({\bf
r},t)     |I>}{<I|I>}
\label{eq:uno}
\end{equation}
and
\begin{equation}
P_{dou}({\bf r},t; \mu , \eta )=\frac{<I|\hat{\psi} ^+_{\mu} ({\bf r},t) \hat{\psi} ^+_{\eta} ({\bf r},t) \hat{\psi}_{\eta} ({\bf r},t) \hat{\psi} _{\mu} ({\bf r},t) |I>}{<I|I>}
\label{eq:dos}
\end{equation}
with $|I>$ the state of the incident beam (expressed in Fock's
space) and where $\mu \vee \eta$ means that we can detect the
particle in any of the two spin states.

The scheme based on the two above equations is the general one to
evaluate detection probabilities in the second quantization
framework, but it is specially well suited to study systems where
the detection occurs through the absorption of the particles to be
measured. As a matter of fact, the method follows closely the
seminal work of Glauber \cite{Gla} to describe the detection of
photons by photoionization. In the usual detection schemes based
on ionization (photographic plates, Geiger counters and
bubblechambers) the detected particle becomes finally mixed with
the rest of components of the detector, i. e., it is absorbed by
the detector.

In the above equations $\hat{\psi}_{\mu} ({\bf r},t)$ is the
Schr\"{o}dinger field operator for spin state $\mu$. Assuming that
we work in a finite volume space, it is given by the expression:
\begin{equation}
\hat{\psi}_{\mu} ({\bf r},t) = \sum _{\bf n} \psi _{{\bf n} \mu}
({\bf r}) \hat{a} _{{\bf n} \mu} (t)
\end{equation}
with $\psi _{{\bf n} \mu}({\bf r})$ a complete set of orthonormal
stationary wavefunctions characterized by the discrete index ${\bf
n}$ (the momenta available to the particle) and spin state $\mu$.
If we would work in $R^3$ instead of the finite volume, the
summation should be replaced by an integration and the discrete
index by a continuous one. The time dependence of
Schr\"{o}dinger's field is contained in the annihilation operator,
which can be expressed as $\hat{a}_{{\bf n} \mu} (t)=\hat{a}_{{\bf
n} \mu} exp(-iE_{ {\bf n} \mu } t/\hbar) $ with $\hat{a}_{{\bf n}
\mu} $ the operator at $t=0$ and $E_{{\bf n} \mu }$ the energy of
the stationary state. The (anti)commutation relations are
$[\hat{a}_{{\bf n} \mu}, \hat{a}^+_{{\bf m} \Omega }]_{\mp}=\delta
^3 _{{\bf n}{\bf m}}\delta _{\mu \Omega}$, with the upper sign of
the double expression valid for bosons and the lower one for
fermions.

The single and double detection events can be seen as two
available channels for the interaction between the incident beam
and the detector. The total probability of detection can be
expressed as:
\begin{equation}
P_{det}({\bf r},t;\mu , \eta )=\alpha _{sin} P_{sin}({\bf
r},t;\mu \vee \eta) + \alpha _{dou} P_{dou}({\bf r},t;\mu , \eta )
\end{equation}
where $\alpha _{sin}$ and $\alpha _{dou}$ are two phenomenological
coefficients characterizing the weight of the two channels. They
can be determined experimentally.

Writing the detection probability in this form we implicitly
assume that the records for a double detection at ${\bf r}$
($P_{dou}({\bf r})$) and for two detections at close points ${\bf
r}$ and ${\bf R}$ inside the detector ($P_{sin}({\bf r})$ and
$P_{sin} ({\bf R})$) can be distinguished. For the Geiger counter
this condition is fulfilled because the spark currents (the
records) are clearly different for both situations. Similarly, in
the case of the bubblechamber for two ionizations we would have
two lines of bubbles (the records), whereas for a double
ionization we would only have one line. The situation is not so
favorable for photographic plates, where the typical size of the
spots (the records) imposes a limit to the possibility of
distinguishing both types of events. However, the typical size of
the spots is very small and the error introduced can be
neglected.

Using the above expressions we can easily evaluate the detection
probabilities for the initial two-particle state
\begin{equation}
|I>=\sum _{{\bf n},{\bf m}} b_{\bf n}d_{\bf m}\hat{a}^+_{{\bf n}
\sigma} \hat{a}^+_{{\bf m} \Omega} |0>
\end{equation}
where ${\bf n}$ and ${\bf m}$ represent the complete set of
momenta available to the particles. On the other hand, $\sigma $
and $\Omega$ are the spin states of the two particles. By
simplicity, we assume that the coefficients representing the
momenta distributions, $b_{\bf n}$ and $d_{\bf m}$, are
independent of the spin states. Finally, $|0>$ represents the
vacuum state.

Using this form of the two-particle incident beam we assume that
it is composed of two particles, one in state
\begin{equation}
\psi _{(b;\sigma )} ({\bf r},t) =\sum _{\bf n} b_{\bf n} \psi
_{{\bf n}\sigma } ({\bf r}) exp(-iE_{\bf n \mu }t/\hbar)
\end{equation}
with $ \psi _{{\bf n}\sigma}$ a complete set of orthonormal
stationary wavefunctions in spin state $\sigma $. The other particle
is in the state $ \psi _{(d;\Omega )} ({\bf r},t)$, given by a
similar expression with obvious modifications.

The denominator of the detection probabilities is
\begin{equation}
<I|I>= \pm 1 + \delta _{\sigma \Omega} |<d|b>|^2
\end{equation}
with $<d|b> =  \sum _{\bf n} d^* _{\bf n} b_{\bf n}$.

A simple calculation gives:
\begin{equation}
P_{sin}({\bf r},t;\mu \vee \eta) = {\cal P} ({\bf r},t;\mu) + {\cal P} ({\bf r},t;\eta)
\end{equation}
with
\begin{eqnarray}
{\cal P} ({\bf r},t; \mu )= \\
\frac{\pm \delta _{\mu \Omega} |\psi _{(d;\Omega )}({\bf r},t)|^2
\pm \delta _{\mu \sigma} |\psi _{(b;\sigma )}({\bf r},t)|^2
+2\delta _{\sigma \Omega} \delta _{\mu \sigma} Re(<d|b> \psi
_{(b;\sigma )}^*({\bf r},t)    \psi _{(d;\Omega )}({\bf
r},t))}{\pm 1+ \delta _{\sigma \Omega} |<d|b>|^2} \nonumber
\end{eqnarray}
and a similar expression for ${\cal P} ({\bf r},t; \eta ) $ with $\mu$ replaced by $\eta $.
On the other hand, we have
\begin{eqnarray}
P_{dou}({\bf r},t;\mu , \eta) = \\
\frac{|\psi _{(d;\Omega )}({\bf r},t)|^2 |\psi _{(b;\sigma )}({\bf
r},t)|^2 (2\delta _{\eta \Omega} \delta _{\mu \sigma} \delta
_{\sigma \eta} \delta _{\Omega \mu} \pm \delta _{\sigma \mu}
\delta _{\eta \Omega} \pm \delta _{\sigma \eta} \delta _{\mu
\Omega}  )}{\pm 1+ \delta _{\sigma \Omega} |<d|b>|^2} \nonumber
\end{eqnarray}
The physical meaning of all these expressions is discussed in next section.

\section{Discussion}

First, we consider the probability of single detections. It must
be noted that both the numerator and the denominator of the
expression for single detection probabilities are negative for
fermions, but the complete expression remains always positive
(these properties can easily be verified following a similar
procedure to that presented in the Appendix of Ref. \cite{San}).

We consider the case ${\cal P} ({\bf r},t; \mu ) $ (the discussion
of ${\cal P} ({\bf r},t; \eta )$ is similar). Three terms
contribute to this single detection probability. Those
proportional to $|\psi _{(d;\Omega)}|^2$ and $|\psi
_{(b;\sigma)}|^2$ represent the probabilities of detecting only
one of the particles. On the other hand, the third term has the
typical form of the interference between the two previous
alternatives for the detection (to detect the particle in state
$(d;\Omega)$ or in $(b;\sigma)$). This last term is only present
when $\sigma = \Omega = \mu$ and $<d|b> \neq 0$. The last
condition implies that both particles must have common modes. When
there are common modes the detector does not distinguish if these
modes correspond to one or the other particle. In the presence of
indistinguishable alternatives quantum theory gives rise to
interference phenomena (see Ref. \cite{San}, where an extensive
analysis of this type of interference has been carried out).

Now, we consider the probability of double detections. Again,
taking into account that $2\delta _{\eta \Omega} \delta _{\mu
\sigma} \delta _{\sigma \eta} \delta _{\Omega \mu} \leq \delta
_{\eta \Omega} \delta _{\mu \sigma}+ \delta _{\sigma \eta} \delta
_{\Omega \mu}$ it is simple to see that the expression is
non-negative for fermions.
This term represents a different type of contribution to the detection
probability. For instance, in the case $\sigma \neq \Omega$, $\mu
= \sigma$, $\eta = \Omega$ and $b=d$ ($b_{\bf n}=d_{\bf n}$) the
total detection probability (valid for both fermions and bosons)
is
\begin{eqnarray}
P_{det}({\bf r},t;\mu , \eta )=\alpha _{sin} (|\psi _{(b;\mu)}
({\bf r},t)|^2 + |\psi _{(b;\eta)} ({\bf r},t)|^2 )+ \\
\nonumber
\alpha _{dou} |\psi _{(b;\mu )}({\bf r},t)|^2 |\psi
_{(b;\eta )}({\bf r},t)|^2
\end{eqnarray}
The second term in the r. h. s. of this expression contains the
product of the squared modulus of the two wavefunctions. This
analytical form differs from the usual squared dependence on the modulus
of the single term.

Other interesting situation is when the incident beam is composed
of two bosons in the same state ($b=d$ and $\mu =\eta = \sigma
=\Omega$). This situation corresponds to a two-boson laser and the
total probability is given by
\begin{equation}
P_{det}({\bf r},t;\mu , \mu )=2\alpha _{sin} |\psi _{(b;\mu)} ({\bf r},t)|^2 + 2\alpha _{dou} |\psi _{(b;\mu )}({\bf r},t)|^4
\end{equation}
The correction to the usual $|\psi |^2$ form of the single-particle
detection distribution is in the form $|\psi |^4$.

We note the similarity between the last result and those obtained
in quantum optics, where the probabilities of one- and two-photon
absorption are, respectively, proportional to the first and second
power of the quantum intensity \cite{Lou}

The results obtained are independent of the mechanism of
detection, ionization or any other, and are valid for any scheme
of detection where various particles can lead to single detection
events (Eqs. (\ref{eq:uno}) and (\ref{eq:dos}) are independent of
the underlying detection mechanism).

\section{Conclusions}

We have analyzed in this Letter a new channel
for the ionization of atoms and molecules by massive particles;
the ionization by interaction with various particles, a massive
counterpart to the well-known phenomenon of multi-photon
absorption. In particular, this phenomenon would lead to {\it
multiple detection events}. The results obtained here would also be
valid for any detection process where the underlying physical
mechanism leads to both types of detections, with one or various
probe particles. The main conclusion is that these events would
give rise to a new term, which when the two particles have the
same mode distribution and spin is proportional to the fourth
power of the modulus of the wavefunction. This distribution can,
in principle, be distinguished from the usual one associated with
single detection events (proportional to the second power of the
modulus of the wave function): if $\alpha _{dou}/\alpha _{sin}$ is
not too small both types of distributions could be distinguished
experimentally. In particular, in atom lasers it seems to be a
realistic goal. The most pictorial manifestation of this behaviour
would be in interferometric arrangements \cite{Adw,Gam}. Placing
the detector at many different locations in a large number of
repetitions of the experiment we could obtain the experimental
curve of the interference experiment. Instead of the usual form
(proportional to $|\psi |^2$) of the experiment with
single-particle beams we would obtain one of the form $\alpha
_{sin} |\psi |^2 + \alpha _{dou} |\psi |^4$, showing a small
correction (of the order $\alpha _{dou}/\alpha _{sin}$) to the
usual pattern.

We have only considered the case of two-particle beams. The
generalization to beams with more particles is straightforward.
The properties of the multi-particle detection can also depend on
the type of incident state: in quantum optics it has been shown
that the absorption efficiency depends on the statistical
properties \cite{pol} and entanglement \cite{Per} of the incident
light.

Most detectors are not sensitive to the spin state of the incident
particles. The results corresponding  to this particular situation
can be obtained directly from our expressions. For incident
identical particles the dependence of the two-particle detection
on the modulus of the wavefunction is in the fourth power, just as
in the example of the previous Section.

The results presented in this Letter show once more the notorious
differences between bosons and fermions. The double sign in all
the expressions for the probability distributions leads to
additions for bosons and subtractions for fermions. The extreme
situation occurs when both incident particles are in very similar
states. Then the bosons tend towards maximum detection
probabilities, whereas the fermion probabilities tend to small
values (in the limit of equal states to null probabilities).


\begin{thebibliography}{99}
\bibitem{com} Geiger counters and bubblechambers are devices too big to enter into
this category of small size detectors. However, a simple trick can overcome this
difficulty. We cover the device with a material impenetrable for the incident beam,
except in a small size window, through which the particles can enter into the detector.
\end{thebibliography}
\end{document}